# Indirect to direct band gap crossover in two-dimensional WS$_{2(1-x)}$Se$_{2x}$ alloys

Cyrine Ernandes[1∥], Lama Khalil[1∥], Hela Almabrouk[1], Debora Pierucci[2], Biyuan Zheng[3], José Avila[4], Pavel Dudin[4], Julien Chaste[1], Fabrice Oehler[1], Marco Pala[1], Federico Bisti[2], Thibault Brulé[5], Emmanuel Lhuillier[6], Anlian Pan[3*], and Abdelkarim Ouerghi[1*]

[1]Université Paris-Saclay, CNRS, Centre de Nanosciences et de Nanotechnologies, 91120, Palaiseau,
[2]CELLS - ALBA Synchrotron Radiation Facility, Carrer de la Llum 2-26, 08290 Cerdanyola del Valles, Barcelona, Spain
[3]Key Laboratory for Micro-Nano Physics and Technology of Hunan Province, State Key Laboratory of Chemo/Biosensing and Chemometrics, and College of Materials Science and Engineering, Hunan University, Changsha, 410082 Hunan, China
[4]Synchrotron-SOLEIL, Université Paris-Saclay, Saint-Aubin, BP48, F91192 Gif sur Yvette, France
[5]HORIBA France SAS, Passage Jobin Yvon, Avenue de la Vauve, 91120 Palaiseau, France
[6]Sorbonne Université, CNRS, Institut des NanoSciences de Paris (INSP) F-75005 Paris, France

[∥]These authors contributed equally to this work.

[*] Corresponding authors E-mail:
anlian.pan@hnu.edu.cn
abdelkarim.ouerghi@c2n.upsaclay.fr

**ABSTRACT**

In atomically thin transition metal dichalcogenide semiconductors, there is a crossover from indirect to direct band gap as the thickness drops to one monolayer, which comes with a fast increase of the photoluminescence signal. Here, we show that for different alloy compositions of WS$_{2(1-x)}$Se$_{2x}$ this trend may be significantly affected by the alloy content and we demonstrate that the sample with the highest Se ratio presents a strongly reduced effect. The highest micro-PL intensity is found for bilayer WS$_{2(1-x)}$Se$_{2x}$ ($x = 0.8$) with a decrease of its maximum value by only a factor of 2 when passing from mono- to bi-layer. To better understand this factor and explore the layer-dependent band structure evolution of WS$_{2(1-x)}$Se$_{2x}$, we performed a nano-angle resolved photoemission spectroscopy study coupled with first-principles calculations. We find that the high micro-PL value for bilayer WS$_{2(1-x)}$Se$_{2x}$ ($x = 0.8$) is due to the overlay of direct and indirect optical transitions. This peculiar high PL intensity in WS$_{2(1-x)}$Se$_{2x}$ opens the way for spectrally tunable light-emitting devices.

**INTRODUCTION**

Continuous improvements in the synthesis of two-dimensional materials have enabled the use of these materials in photonics, opto-electronics as well as the emergence of advanced nano-technologies[1,2]. Particularly, transition metal dichalcogenides (TMDs), which possess a high carrier mobility[2] and several interesting spin properties[3–5] constitute promising candidates for the study of emergent physical phenomena and functionalities in electronics[5], photonics and superconductivity[6]. The successful manipulation of these properties makes it possible to exploit the full potential of the TMDs, such as the synthesis of semiconductor alloys with tunable band gaps obtained through the change of chemical compositions[7–10] and their integration into catalytic[11,12], opto-electronic and energy related devices[13]. More interestingly, stacked structures of TMDs can improve device performances leading to original device concepts[14]. As examples, multilayer and vdW heterostructures were notably integrated in vertical tunneling FETs and LEDs[15]. The physical characteristics of those 2D materials (bandgap, carrier mobility, doping…)[16] can be tailored by controlling the number of layers (i.e., the thickness)[17,18], the stoichiometric composition, the strain[19–22], the dielectric or electric field environments[23]. There is a very active competitive field, mainly driven by exfoliation, to achieve bilayer or multilayer of 2D material, especially for superconductivity but also for optics. Synthesis techniques such as chemical vapor deposition (CVD) give access to large scale well-controlled bilayer structures with on-demand properties, like a controlled angle between the two lattices. For many reasons, 2D materials, where the optical response remains high in a bilayer or multilayer configurations, are highly sought after.

Furthermore, numerous studies on particular TMDs ($MoS_2$, $MoSe_2$, $WSe_2$, $WS_2$) have revealed that they present a tunable band gap that transits from the indirect to direct band gap when these materials are reduced to a single layer thin film accompanied by a significant improvement in the photoluminescence PL intensity (an increase by one or two order of magnitude)[24]. However, this property which was considered as a common point of all the semiconducting TMDs does not seem to be applicable for certain materials such as $MoTe_2$[25]. Indeed, recent works have shown that the indirect-to-direct band gap crossover occurs in the $MoTe_2$ trilayer thus showing that the single and bi-layer being direct bandgap semiconductors. In particular, it was demonstrated that $MoTe_2$ bilayer still maintains a high PL intensity, which is only 2-3 times weaker than that of the single layer.

In this paper, few-layers of different alloy compositions of $WS_{2(1-x)}Se_{2x}$ have been studied in order to uncover the indirect to direct gap crossover as a function of the number of layers. To this purpose, the optical band gap evolution of few-layer alloys was studied using µ-PL measurements and compared to few thin-layers of $WS_2$ and $WSe_2$. We establish that the measured photoluminescence in $WS_{2(1-x)}Se_{2x}$ follows the same trend as that observed in $WSe_2$ and $WS_2$ for the alloy compositions where $x$ = 0.3, 0.5 and 0.8. However, the µ-PL associated to the bilayer thickness for the alloys with higher Se content ($x$ = 0.8) still exhibits a high intensity compared to the monolayer (ML), similarly to thin-layer semiconducting $MoTe_2$. To better understand this phenomenon and its origin, the electronic properties have been studied using nano-ARPES and DFT calculations. Our results directly demonstrate the presence of a direct band gap for monolayer alloys and of an indirect band gap for the increasing number of layers. Therefore, the $WS_{2(1-x)}Se_{2x}$ alloy monolayer is a direct-gap semiconductor, while the $WS_{2(1-x)}Se_{2x}$ bilayer presents an overlay of direct and indirect optical transitions. This robustness of PL intensity will ease material integration because it is more tolerant to thickness fluctuation.

## RESULTS AND DISCUSSION

### Band gap tunability

Using CVD (see the Methods section), we synthesized few-layer $WS_{2(1-x)}Se_{2x}$ samples with different chemical compositions (where $x$ = 0.3, 0.5 and 0.8) on $SiO_2$/Si substrates[26]. The resulting samples are typically made of single-crystal domains with well-defined hexagonal or triangular shapes[27]. Because $WS_2$ and $WSe_2$ share same crystal structure, the random $WS_{2(1-x)}Se_{2x}$ alloy can be simply represented as in Fig. 1a where the chalcogen atom site can be either occupied by S or Se (within their relative abundance), without any other distortion or super-periodicity. In Fig. 1b, we show an optical image of a representative flake, identified from its darker optical contrast with respect to the underlying substrate, exhibiting an equilateral triangle shape.

To study the band gap tunability associated to the modification of the S/Se stoichiometric ratio in the WSSe layer[28], we performed µ-PL measurements, using a confocal microscope setup[29]. We recorded PL spectra for each alloy on mono-, bi- and tri-layer (1, 2 and 3 ML) thick flakes, which we compared to the ones of pure $WS_2$ and $WSe_2$ samples. As shown in Fig. 1c and Supplementary Fig. 1, the PL spectral peaks of the single-layer $WS_{2(1-x)}Se_{2x}$ crystals with a Se composition at 0.8, 0.5 and 0.3 are located at 1.70, 1.77 and 1.81 eV, respectively[28]. Their intermediate position between the single-layer $WSe_2$ and $WS_2$ peaks (located at 1.63 eV and 1.95 eV, respectively) confirms that the band gap can be continuously adjusted by alloying two TMDs. The absence of any signal at the location of the pure compounds (1.63 or 1.95 eV) and the conservation of the PL line width along the compositional range, are strong indications for a complete intermixing of the two different chalcogen atoms in the alloy (i.e., random alloy), excluding the presence of separate $WS_2$ or $WSe_2$ domains[30,29]. Upon increasing the number of layers, the PL peak shifts to lower energies and decreases in intensity for all S/Se ratios, see Supplementary Fig. 2. In contrast to the findings for $Mo_{(1-x)}W_xS_2$[31], we do not see any bowing of the band gap as a function of $x$ for $WS_{2(1-x)}Se_{2x}$. Our results are in agreement with a previous PL study on the $WS_{2(1-x)}Se_{2x}$ alloy[28]. Note that these are very different materials, $Mo_{(1-x)}W_xS_2$ is obtained by varying the metal (W,Mo) composition, while the present $WS_{2(1-x)}Se_{2x}$ alloys only differ by the chalcogen fraction (S, Se). This has strong implications in the respective composition of the HOMO (highest occupied molecular orbital) and LUMO (lowest unoccupied molecular orbitals) orbitals. In the case of the metal alloying (W, Mo), the contribution of metal elements to the HOMO is identical for $MoS_2$ and $WS_2$ ($d_{xy}$ and $d_{x2-y2}$ orbitals), while the LUMO changes from $d_{z2}$ orbital in $MoS_2$ to $d_{xy}$, $d_{x2-y2}$, and $d_{z2}$ orbitals in $WS_2$. In the case of chalcogen alloying (Se, S), the orbital contributions to both the HOMO and LUMO of $WS_2$ and $WSe_2$ are identical. The HOMO is composed of the $d_{xy}$ and $d_{x2-y2}$ orbitals of W, while the LUMO contains the $d_{xy}$ and $d_{x2-y2}$ orbitals of W and the $p_x$, $p_y$ orbitals of S and Se. These matching contributions explain the simple linear dependence of the bandgap as a function of the chalcogen composition in $WS_{2(1-x)}Se_{2x}$ as measured by PL measurements and confirmed by DFT simulations (see Supplementary Fig. 3).

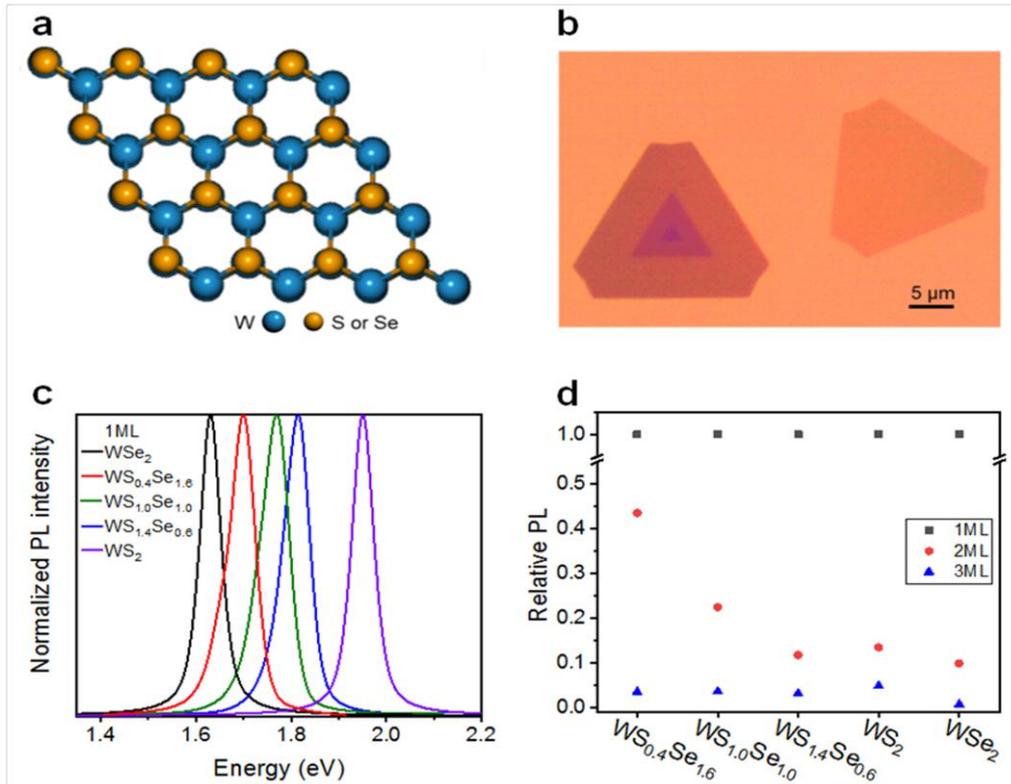

**Fig. 1: Structural and electronic properties of the few layers WS$_{2(1-x)}$Se$_{2x}$. a** Schematic views of the atomic structure of the WS$_{2(1-x)}$Se$_{2x}$. For clarity, only two superposed atomic sheets are displayed, in which the atoms in the top layer have been drawn slightly smaller than the ones in the bottom layer. **b** Typical optical image of the as-grown WS$_{2(1-x)}$Se$_{2x}$ flakes on the SiO$_2$/Si substrate. **c** Normalized PL spectra from monolayer WS$_2$, WSe$_2$ and different alloy compositions of monolayer WSSe. **d** The evolution of the relative PL intensity of mono-, bi- and tri-layers of WS$_2$, WSe$_2$ and WSSe. All measurements were recorded at room temperature with a 532 nm laser excitation.

**Atypical behavior of WS$_{2(1-x)}$Se$_{2x}$ ($x = 0.8$)**

To compare PL signals originating from different layer thicknesses at given chemical composition, we normalize all the data to the maximum PL intensity of the 1 ML sample, see Fig. 1d. The WS$_{2(1-x)}$Se$_{2x}$ ($x = 0.3$) exhibits a similar behavior as WSe$_2$ and WS$_2$: by increasing the layer thickness from mono- to bilayer, the maximum PL intensity decreases by one order of magnitude[26]. For higher Se contents $x = 0.5$ ($x = 0.8$, respectively), the 2 ML maximum PL intensity is reduced by a factor of 5 (2, respectively) with respect to the 1 ML sample. Note that a similar low factor of 2 has been previously observed in MoTe$_2$ by I. Lezama *et al.*[25], in which the authors concluded that bilayer MoTe$_2$ is a direct band gap semiconductor. Another important peculiarity of WS$_{2(1-x)}$Se$_{2x}$ ($x = 0.8$) is the PL broadening increase going from 1ML (69 meV FWHM) to 2ML (92 meV FWHM), see Supplementary Fig. 2. This indicates the presence of additional radiative contributions to PL spectrum, other than the direct band transition.

Similar attenuated drop of the maximum PL intensity by only a factor of 2 when passing from mono- to bi-layer of WS$_{2(1-x)}$Se$_{2x}$ ($x = 0.8$), can be traced on two distinct flakes presenting 2H and 3R stackings[32]. In Fig. 2a, e, we present the optical images, obtained from a triangular and another hexagonal flake. The change in contrast represents different thicknesses of the flake. Their corresponding PL intensity and peak position maps are reported in Fig. 2b, c for the triangular flake and in Fig. 2f, g for the hexagonal one, respectively. The changing in the intensity contrast in the PL intensity maps confirms that the number of layers varies across each flake: a higher intensity scale actually indicates a decrease in the number of layers and vice versa[29]. This is also proved by taking punctual PL spectra in each region of each flake, see Fig. 2d, h. By comparing the spectra of Fig. 2d, h to the ones of Supplementary Fig. 2e, obtained from another sample with the same S/Se stoichiometric ratio, we find a high degree of reproducibility of our results. Thus, the observed peculiar behavior of the sample with the Se content at 80% does not depend on the shape of the flake or on its stacking order[16]. Note that the PL intensity and peak position maps show a high contrast and composition uniformity for each thickness, which indicates that the optical properties are homogeneous in each region and that the samples are of high crystalline purity. The micro-Raman data also confirm that, in the limits of the experimental resolution of our setup (around 0.5 µm), the pure $WS_2$/$WSe_2$ phases are absent, see Supplementary Fig. 4.

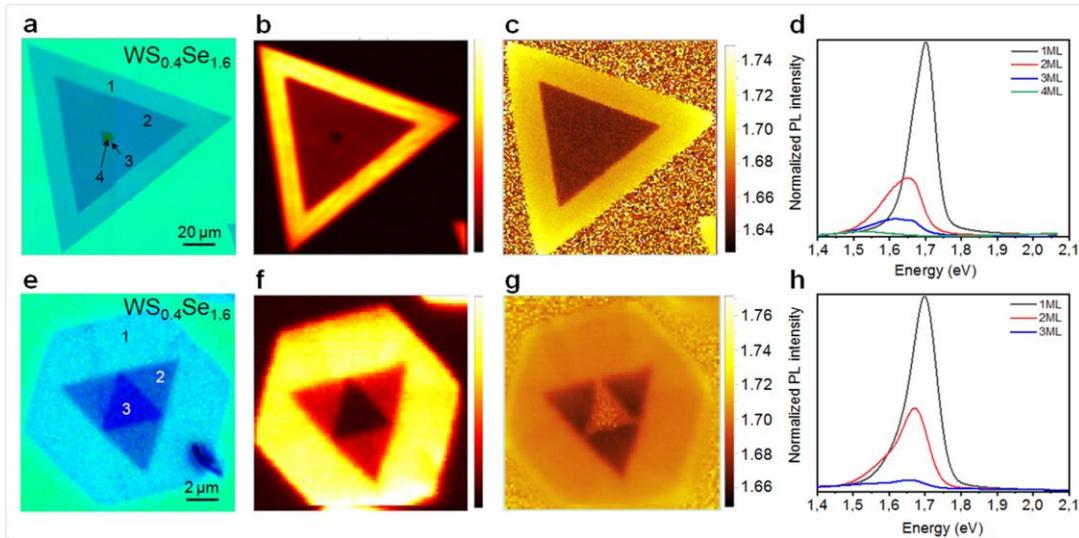

**Fig. 2: Optical properties of the $WS_{2(1-x)}Se_{2x}$ ($x = 0.8$) sample. a, e** Optical images obtained from two distinct triangular and hexagonal flakes, showing that the 1, 2 and 3 ML regions present different optical contrasts. **b, f** PL intensity maps and **c, g** peak position maps acquired on the triangular and hexagonal flakes of Fig. 2a, e with a 532 nm laser excitation. **d, h** Typical PL spectra obtained from different positions on the flakes of Fig. 2a, e indicated by 1, 2, 3 and 4 in (**a**), corresponding to 1, 2, 3 and 4 ML, respectively.

**Evidence for an indirect-to-direct band gap crossover**

Further insight into the electronic properties of $WS_{2(1-x)}Se_{2x}$ ($x = 0.8$) can be obtained by performing nano-XPS/nano-ARPES measurements at low temperature ($T_{sample}$ = 70 K)[33,34]. The precise electronic characterization of the distinct regions on the flakes becomes possible by adding spatial resolution to ARPES experiments[33]. The CVD-grown flakes were transferred onto epitaxial graphene on SiC before conducting the nano-XPS/ARPES measurements to avoid charging effects[35–38], see the Methods section. By integrating the photoemission intensity within a selected energy window around the Se *3d* peak, while scanning the sample along two in-plane directions, we can first get morphological information on the sample, see Fig. 3a. Red-, orange- and yellow-colored regions correspond to the 1, 2 and 3 ML thicknesses, respectively, and the blue-colored regions are those of the substrate. Each of these regions display a very uniform intensity, attributed to the homogeneity of the electronic properties. Note that in the limits of our spatial resolution, we do not detect any small domains of $WS_2$ or $WSe_2$, related to the presence of structural defects and grain boundaries. The left (right) panel of Fig. 3b shows three µ-XPS spectra of the Se *3d* (W *4f*, respectively) core level acquired on three distinct points of each region. The experimental data points are represented as black, red and blue circles, the Voigt envelop of fitted components is shown as a green solid line, and the fitting components are plotted as orange and blue curves[30]. All the Se *3d* spectra consist of a single pair of spin-orbit doublet, corresponding to Se *3d*$_{5/2}$ and Se *3d*$_{3/2}$ and attributed to the bonding with W (spin-orbit splitting of 0.83 eV, *3d*$_{3/2}$:*3d*$_{5/2}$ ratio of 0.6, and FWHM of 0.5 eV). The W *4f* spectra (Fig. 3c), on the other hand, can be well fitted by two pair of spin-orbit doublets, corresponding to W *4f*$_{5/2}$ and W *4f*$_{7/2}$ and relative to the bonding with the two chalcogen atoms S and Se (spin-orbit splitting of 2.15 eV, *4f*$_{5/2}$:*4f*$_{7/2}$ ratio of 0.7, and FWHM of 0.4 eV)[39]. The ratio between the area of the S-related component is about 20% of the total one (the sum of the components related to S and Se), which reflects the chemical composition of the alloy. No additional components are present in the nano-XPS spectrum confirming the homogeneity of the sample and excluding the presence of possible covalent bonds between the WSSe and the graphene substrate, as expected in a sharp van der Waals WSSe/graphene interface. Moreover, the Se *3d* and W *4f* energy positions vary according to the region. This shift can be related to the band alignment at the interface between graphene and WSSe[40,41], in agreement with the nano-ARPES results discussed in the coming paragraph. These photoemission spectra are used to estimate the Se contents of the WSSe alloys. The total amount of Se atoms is calculated from the Se *3d* and W *4f* peak area ratio in the surface sensitive configuration, weighted by their relative cross sections.

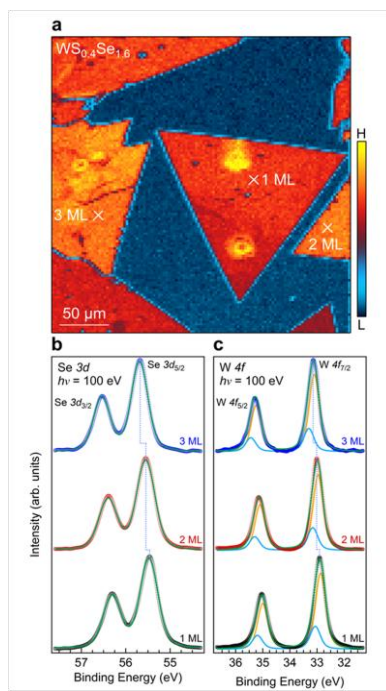

**Fig. 3: Nano-XPS maps of the $WS_{2(1-x)}Se_{2x}$ ($x$ = 0.8) sample. a** Spatially-resolved photoemission map (of 300 × 300 µm² area) of the Se *3d* core level obtained from the $WS_{2(1-x)}Se_{2x}$ ($x$ = 0.8) sample by using a photon energy of 100 eV. **b**, **c** Left (Right) panel corresponds to the µ-XPS spectra of the Se *3d* (W *4f*) core level acquired on three distinct points represented as crosses in panel (**a**).

In order to unambiguously evaluate the direct or indirect bandgap character of the 1ML, 2ML and 3ML of WS$_{2(1-x)}$Se$_{2x}$ ($x = 0.8$), we investigated the electronic band structure of this alloy crystal experimentally by means of ARPES and theoretically with DFT calculations[42,43]. Figures 4a-c present the photoemission intensity maps acquired at 100 eV on mono, bi- and tri-layers of WS$_{2(1-x)}$Se$_{2x}$ ($x = 0.8$), respectively along the Γ-K direction[44]. Their second derivatives on which the theory (blue dashed lines) is overplotted are instead reported in Fig. 4d-f. The top of the valence band at the K point is mostly formed by planar $d_{xy}$ and $d_{x2-y2}$ orbitals of tungsten, while at the Γ point the band is mostly composed by W $d_z$ orbitals and S, Se $p_z$ orbitals based on DFT calculations[30]. Because of their out-of-plane character, the bands with the lowest energy at Γ are the most sensitive to the number of layers composing the system and, as shown in Fig. 4, the single down-dispersing parabola of the 1ML splits into two (three) parabolas for 2ML (3ML)[45]. This detailed evolution of the valence band structure for different WS$_{2(1-x)}$Se$_{2x}$ ($x = 0.8$) thicknesses is highly useful in terms of offering a direct way to determine the WS$_{2(1-x)}$Se$_{2x}$ ML number.

For 1ML, the valence band maximum (VBM) is located at the K point at 1.16 eV binding energy (BE), while for 2 and 3 ML it is positioned at the Γ point at 1.16 eV and 1.27 eV BE, respectively. Consequently, by increasing the number of ML, the WS$_{2(1-x)}$Se$_{2x}$ ($x = 0.8$) quasiparticle band gap decreases, inducing also a shift in the VBM position. Going from Γ to K, the character of the valence band at the top changes from out-plane to in-plane. Thus, the split due to the number of layers is removed in favour of only a spin-orbit coupling split at the K point of the in-plane W $d$ orbitals. The values of the spin orbit coupling (SOC) are of 500, 420 and 380 meV for mono-, bi- and tri-layer WSSe, respectively, which represent high values compared to other 2D materials[45]. Besides, the energy difference between the K and Γ points ($E_K - E_\Gamma$) is about 500 meV, -60 meV and -480 meV for mono-, bi- and tri-layer WSSe, respectively. A similar trend was already observed for few layer WSe$_2$ by Y. Zhang et al.[46] In their case, the energy difference between the K and Γ points at the VBM ($E_K - E_\Gamma$) was about 560 meV, -80 meV and -11 meV for mono-, bi- and tri-layer WSe$_2$, respectively. For bi-layer WS$_{2(1-x)}$Se$_{2x}$ ($x = 0.8$), the fact that the top of the VB at K point is slightly lower than the one at Γ suggests that 2 ML of WS$_{2(1-x)}$Se$_{2x}$ ($x = 0.8$) may show a significant contribution from both direct and indirect emissions in the final PL spectrum[47]. In Fig. 4d-f, the second derivatives of the experimental data were compared with the band structure calculations in the DFT framework for freestanding 2H-WS$_{2(1-x)}$Se$_{2x}$ ($x = 0.8$) layers. The DFT calculated bands are shifted to account for the Fermi level position. The main features are well reproduced by the calculated band structures along the ΓK high symmetry direction for all thicknesses. Both the measured and DFT calculated band structures agree for the 2H ordering of WS$_{2(1-x)}$Se$_{2x}$ ($x = 0.8$). As expected, 1ML WSSe is a direct-gap semiconductor (at the K point). On the other hand, 3ML WSSe behaves differently, showing a VBM located at Γ and a conduction band minimum (CBM) located at K. Consistently with the ARPES image of bilayer WSSe (Fig. 4b), the local maximum of the VB at K is very close to the one at Γ. Similarly, the local minimum of the CB at the Λ point (the middle-point between K and Γ) is very close to the CBM at K. This is a consequence of the increased hopping between the $p_z$ orbitals of the chalcogen atoms in multi-layer TMDs, which tends to increase the top of the VB at Γ and decrease the minimum of the CB at Λ. From this result we can deduce that bilayer WS$_{2(1-x)}$Se$_{2x}$ ($x = 0.8$) is a semiconductor where direct and indirect optical transitions can both contribute to the PL spectra. The additional contribution of these two transitions explains the relatively large PL intensity of 2ML with respect to 1ML WS$_{2(1-x)}$Se$_{2x}$ ($x = 0.8$).

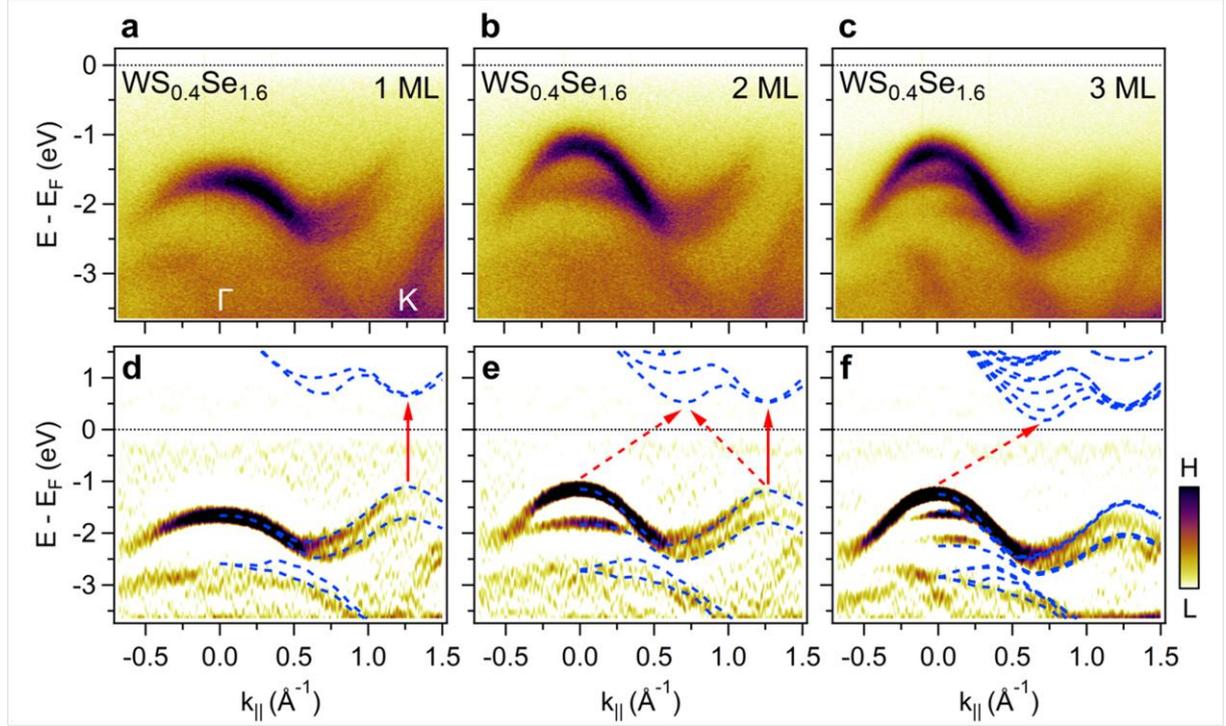

**Fig. 4: Electronic band structure for mono and few layers of $WS_{2(1-x)}Se_{2x}$ ($x = 0.8$).** Nano-ARPES images acquired on: **a** 1 ML, **b** 2 ML, and **c** 3 ML $WS_{2(1-x)}Se_{2x}$ ($x = 0.8$) along the Γ-K direction and taken at a photon energy of 100 eV. **d-f** Second derivative spectra of the (**a-c**) panels on which the calculated band structure obtained by DFT (blue dashed lines) is superimposed.

In summary, we have successfully studied the electronic properties of different thicknesses of $WS_{2(1-x)}Se_{2x}$ alloys with tunable content on sulfur and selenium. We demonstrated that the PL intensity increases when the number of layer decreases. Remarkably, we found that for the high Se content, $WS_{2(1-x)}Se_{2x}$ ($x = 0.8$) exhibits an atypical behavior when the number of layers is reduced. For this particular composition, we observe a high PL signal for both monolayer and bilayer with a drop of the maximum intensity by only a factor 2 when switching from mono- to bilayers. This is interpreted as the signature of the presence of competing contributions of direct and indirect optical transitions, which was confirmed experimentally by ARPES and theoretically by means of DFT calculations. Consequently, by changing the alloy content and the thickness, one can achieve spectrally tunable and robust electroluminescent devices.

**METHODS**

**Growth of $WS_{2(1-x)}Se_{2x}/SiO_2/Si(001)$:** $WS_{2(1-x)}Se_{2x}$ samples were grown by chemical vapor deposition (CVD) in a 1" quartz tube furnace. A quartz boat contained $WS_2$ and $WSe_2$ powder was loaded at the center of the furnace. $SiO_2/Si$ substrate was placed at the downstream of the furnace. Before heating, an Ar gas flow was introduced into the system for 10 min in order to exhaust the air and maintain the flow at 70 standard-state cubic centimeter per minute (sccm). The furnace was then rapidly heated to 1100 °C in 30 min. After keeping the growth at this temperature for 5 minutes, the furnace was then cooled down to room temperature naturally.

**μ-PL measurements:** The μ-PL measurements were conducted at room temperature, using a commercial confocal Horiba micro-Raman microscope with a 100× objective and a 532 nm laser excitation. The laser beam was focused onto a small spot having a diameter of ~1 μm on the sample and its incident power was about 5 μW. All measurements are acquired with a 532 nm laser excitation at room temperature.

**Band structure of few layers $WS_{2(1-x)}Se_{2x}$ :** The ARPES measurements were conducted at the ANTARES beamline of Synchrotron SOLEIL (Saint-Aubin, France). We used linearly horizontal polarized photons of 100 eV and a hemispherical electron analyzer with vertically-confining entrance slit to allow band mapping. The total angle and energy resolutions were 0.25° and 10 meV. All nano-XPS/ARPES experiments were done at low temperature (70 K).

**Theoretical calculations:** we performed plane-wave density functional theory (DFT) simulations by means of the Quantum EPSRESSO code[48]. To include the spin-orbit interaction, we realized noncollinear calculations with the adoption of fully relativistic norm-conserving pseudopotentials. In order to attain a better estimation of the band gap, we adopted the HSE hybrid functional[49] to approximate the exchange-correlation term. The self-consistent solution was obtained by using a 15x15x1 Monkhorst-Pack k-points grid centered around the Γ point and a cutoff energy of 50 Ry. In layered materials, the van der Waals forces were accounted by means of the semiempirical Grimme DFT-D3 correction[50]. A vacuum space of 24 Å in the *z* direction was assumed in the unit cell to suppress the interaction between two adjacent sheets in the periodic arrangement. The cell parameters and atomic coordinates were fully relaxed by using a convergence threshold for forces and energy of $10^{-3}$ and $10^{-4}$ (a.u,), respectively. The WSSe was simulated within the virtual crystal approximation (VCA), where at each chalcogen position we located a virtual atom whose pseudopotential is given by a linear interpolation of the pseudopotentials for S and Se. Such a mean-field approximation is expected to provide the correct trends of the band-structure evolution of samples with sufficiently large surfaces.

## DATA AVAILABILITY

The datasets generated during and/or analyzed during the current study are available from the corresponding author on reasonable request.

## ACKNOWLEDGMENTS

We acknowledge the financial support by RhomboG (ANR-17-CE24-0030), MagicValley (ANR-18-CE24-0007) and Graskop (ANR-19-CE09-0026) grants. This work is supported by a public grant overseen by the French National Research Agency (ANR) as part of the "Investissements d'Avenir" program (Labex NanoSaclay, ANR-10-LABX-0035). Support from the Materials Engineering and Processing program of the National Science Foundation, Award Number CMMI 1538127 is greatly appreciated. Support from the National Natural Science Foundation of China (51525202 and U19A2090).

## AUTHOR CONTRIBUTIONS

B. Z. and A. P. fabricated the samples. C. E., L. K., D. P., A. O., J. A. and P.D. carried out the nano-XPS/nano-ARPES experiments. J. C., A. O. and T. B. characterized the samples by means of μ-Raman/PL spectroscopy and analyzed the Raman/PL data. M. P. carried the DFT calculation. C. E., J. C., L. K., J. A., D. P., F. B., F. O.,

A. O, and E. L. analyzed the data. C. E., M. P. and A. O. wrote the manuscript. All the authors discussed the results and commented on the manuscript.

# Supplementary Information

## Indirect to direct band gap crossover in two-dimensional WS$_{2(1-x)}$Se$_{2x}$ alloys

Cyrine Ernandes[1,7], Lama Khalil[1,7], Hela Almabrouk[1], Debora Pierucci[2], Biyuan Zheng[3], José Avila[4], Pavel Dudin[4], Julien Chaste[1], Fabrice Oehler[1], Marco Pala[1], Federico Bisti[2], Thibault Brulé[5], Emmanuel Lhuillier[6], Anlian Pan[3*], and Abdelkarim Ouerghi[1*]

[1]Université Paris-Saclay, CNRS, Centre de Nanosciences et de Nanotechnologies, 91120, Palaiseau,
[2]CELLS - ALBA Synchrotron Radiation Facility, Carrer de la Llum 2-26, 08290 Cerdanyola del Valles, Barcelona, Spain
[3]Key Laboratory for Micro-Nano Physics and Technology of Hunan Province, State Key Laboratory of Chemo/Biosensing and Chemometrics, and College of Materials Science and Engineering, Hunan University, Changsha, 410082 Hunan, China
[4]Synchrotron-SOLEIL, Université Paris-Saclay, Saint-Aubin, BP48, F91192 Gif sur Yvette, France
[5]HORIBA France SAS, Passage Jobin Yvon, Avenue de la Vauve, 91120 Palaiseau, France
[6]Sorbonne Université, CNRS, Institut des NanoSciences de Paris (INSP) F-75005 Paris, France
[7]These authors contributed equally to this work: Cyrine Ernandes, Lama Khalil.

[*] Corresponding authors E-mail:
anlian.pan@hnu.edu.cn
abdelkarim.ouerghi@c2n.upsaclay.fr

**Supplementary Figures**

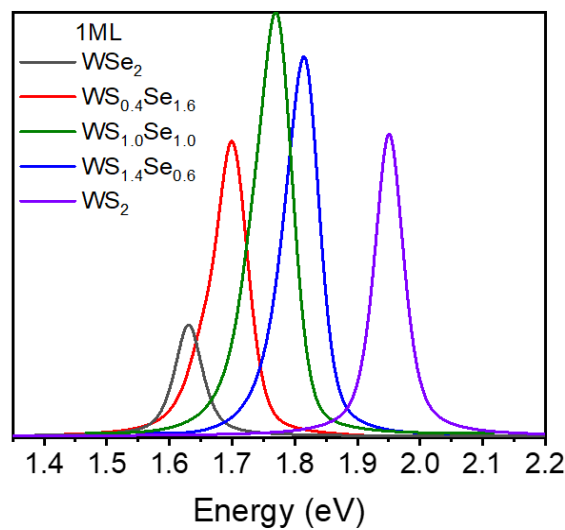

**Supplementary Figure 1. Photoluminescence study of $WS_{2(1-x)}Se_{2x}$ ($x$ = 0.3, 0.5 and 0.8), $WS_2$ and $WSe_2$.** PL spectra from monolayer $WS_2$, $WSe_2$ and different alloy compositions of monolayer WSSe.

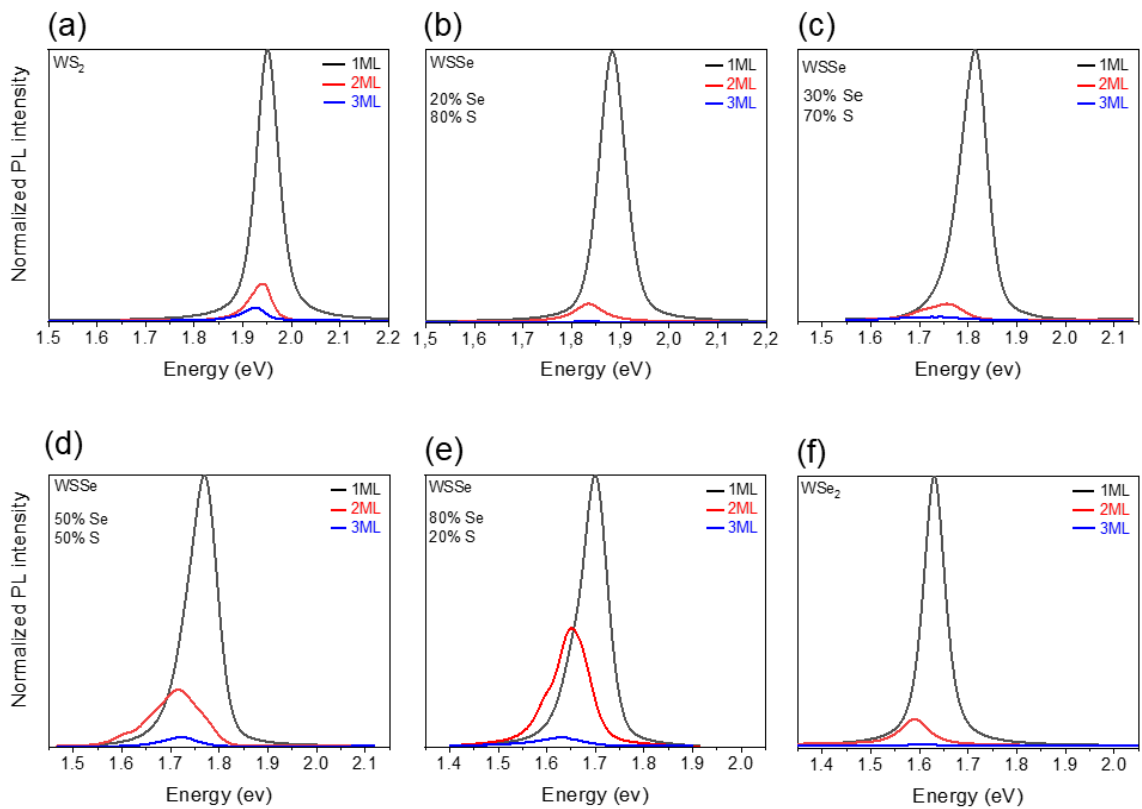

**Supplementary Figure 2. Evolution of the photoluminescence (PL) spectra as a function of the chemical composition.** PL spectra of mono- (black), bi- (red) and trilayers (blue) of: **a** $WS_2$, **b-e** different $WS_{2(1-x)}Se_{2x}$ ($x$ = 0.3, 0.5 and 0.8) alloys and **f** $WSe_2$. All measurements were recorded at room temperature and with a 532 nm laser excitation.

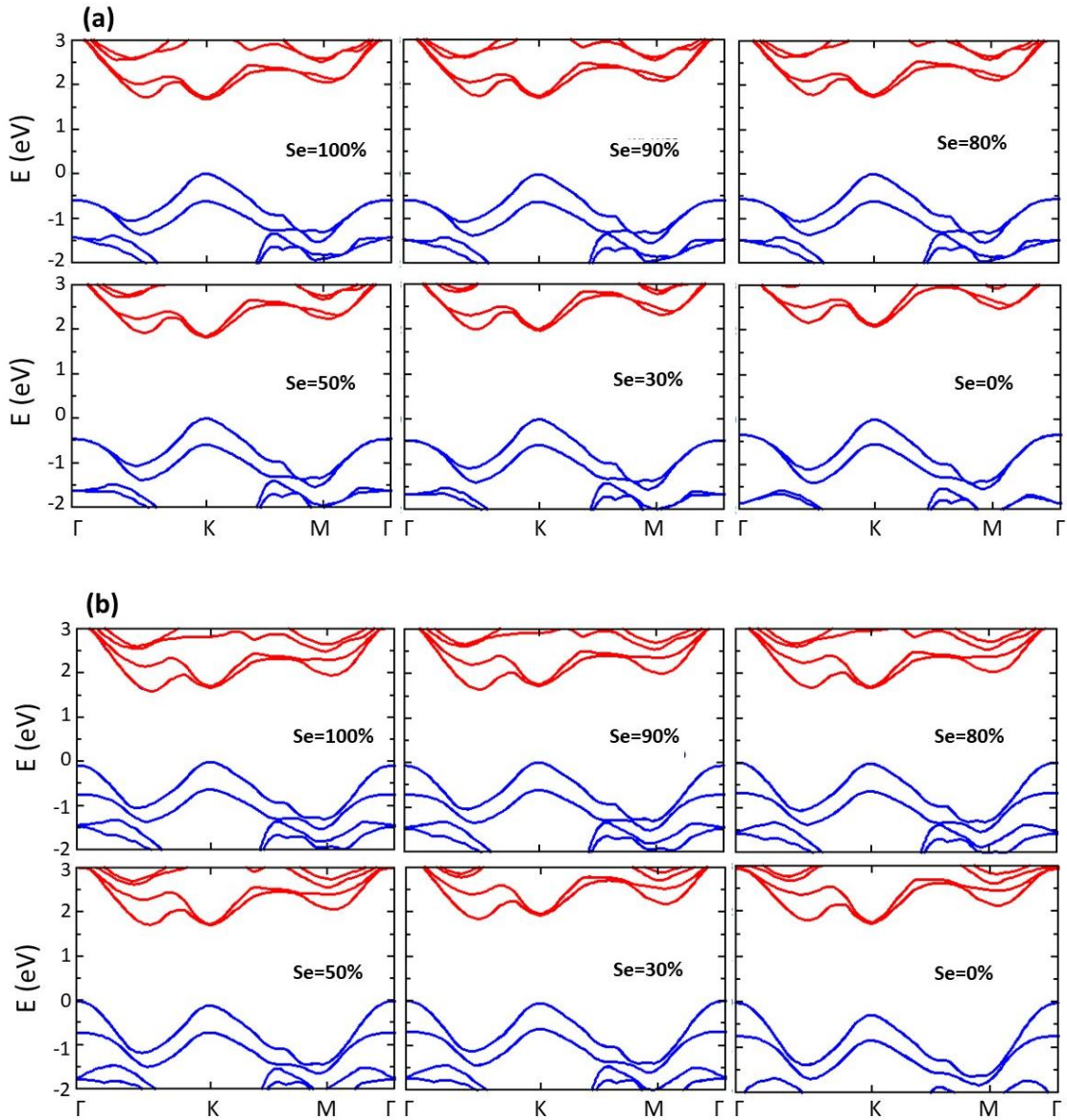

**Supplementary Figure 3. Band structures of the WS$_{2(1-x)}$Se$_{2x}$ films. a** This panel displays the electronic band structures along high-symmetry directions of monolayer WS$_{2(1-x)}$Se$_{2x}$ for $x$ = 1, 0.9, 0.8, 0.5, 0.3 and 0 computed within the DFT framework and an hybrid functional HSE to approximate the exchange-correlation potential. **b** This panel shows the same quantities for the bilayer WS$_{2(1-x)}$Se$_{2x}$. A roughly linear dependence on the Se concentration is found for the bandgap of both 1ML and 2ML. Blue (red) lines correspond to valence (conduction) bands.

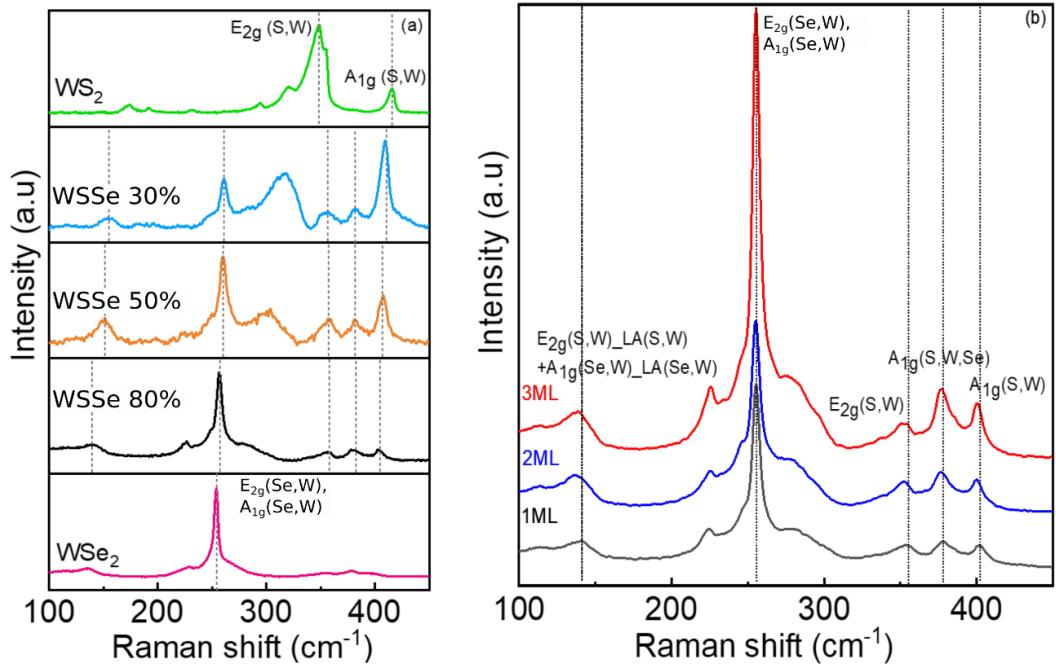

**Supplementary Figure 4. Micro-Raman characterization of the $WS_{2(1-x)}Se_{2x}$ alloys. a** Comparison between the micro-Raman modes of single layer $WS_{2(1-x)}Se_{2x}$ ($x$ = 0.3, 0.5 and 0.8) with the $WS_2$ and $WSe_2$ ones. All spectra were acquired at room temperature. The Raman spectra of the $WS_{2(1-x)}Se_{2x}$ alloys (with different S/Se stoichiometric ratios) all present the same five peaks, which correspond to the same modes but which differ in position and intensity according to the constitution of the alloy. The Raman spectrum of $WS_2$ consists of many first and second order peaks. In particular, the $A_{1g}(S,W)$ mode represents the out of plane vibrations of the sulfur atoms, which shows a resonance peak at 416.9 cm$^{-1}$. For $WS_{2(1-x)}Se_{2x}$, this peak shifts to the lower frequencies with a decreasing intensity when more selenium is introduced. $WS_2$ also shows a higher resonance peak, namely the $E_{2g}(S,W)$ mode, at 349 cm$^{-1}$. This mode represents the in plane vibrations of tungsten and shifts towards higher frequencies with a decrease in its intensity when the Se content increases in $WS_{2(1-x)}Se_{2x}$. Note that in this Raman study we only focus on these two modes for $WS_2$. Similarly, the Raman spectrum of $WSe_2$ consists of many peaks, but we focus on the main one: the $A_{1g}(Se,W)$ mode. This resonance peak is located at 253.9 cm$^{-1}$ and is nearly at the same frequency with respect to the 1 ML $WS_{2(1-x)}Se_{2x}$ alloy peaks. **b** micro-Raman spectra of $WS_{2(1-x)}Se_{2x}$ ($x$ = 0.8) for the thicknesses ranging from 1 ML to 3 ML. The spectra exhibit five modes, which can be assigned to the $A_{1g}(S,W)$ (at ~400 cm$^{-1}$), $A_{1g}$ (S,W,Se) (at ~377.8 cm$^{-1}$), $E_{2g}(S,W)$ (at ~353.5 cm$^{-1}$), $E_{2g}(Se,W)$ and $A_{1g}(Se,W)$ (both at ~255 cm$^{-1}$), and $E_{2g}(S,W)\_LA(S,W)+A_{1g}(Se,W)\_LA(Se,W)$ (at ~138.5 cm$^{-1}$) modes. Note that the position of the $A_{1g}(S,W)$, $A_{1g}$ (S,W,Se) and $E_{2g}(S,W)$ peaks shifts towards higher frequencies and shows an intensity evolution when the number of layer is decreased. This allowed us to determine the layer thickness in our samples. Two peaks can only be observed for the $WS_{2(1-x)}Se_{2x}$ alloys and not for $WS_2$ and $WSe_2$: the $E_{2g}(S,W)\_LA(S,W)+A_{1g}(Se,W)\_LA(Se,W)$ mode, which represents the superposition of $E_{2g}(S,W)\_LA(S,W)$ and $A_{1g}(Se,W)\_LA(Se,W)$ modes, and the $A_{1g}(S-W-Se)$ one.